# Node.DPWS – High performance & scalable Web Services for the IoT

**Konstantinos Fysarakis**

Dept. of Electronic & Computer Engineering

Technical University of Crete

Chania, Crete, Greece

kfysarakis@isc.tuc.gr

**Damianos Mylonakis**

Dept. of Computer Science

University of Crete

Heraklion, Crete, Greece

i@danmilon.me

**Charalampos Manifavas**

Dept. of Informatics Engineering

Technological Educational Institute of Crete

Heraklion, Crete, Greece

harryman@ie.teicrete.gr

**Ioannis Papaefstathiou**

Dept. of Electronic & Computer Engineering

Technical University of Crete

Chania, Crete, Greece

ypg@mhl.tuc.gr

**Abstract.** *Interconnected computing systems, in various forms, are expected to permeate our lives, realizing the vision of the Internet of Things (IoT) and allowing us to enjoy novel, enhanced services that promise to improve our everyday lives. Nevertheless, this new reality also introduces significant challenges in terms of performance, scaling, usability and interoperability. Leveraging the benefits of Service Oriented Architectures (SOAs) can help alleviate many of the issues that developers, implementers and end-users have to face in the context of the IoT. This work presents Node.DPWS, a novel implementation of the Devices Profile for Web Services (DPWS) based on the Node.js platform. Node.DPWS can be used to deploy lightweight, efficient and scalable Web Services over heterogeneous nodes, including devices with limited resources. The performance of the presented work is evaluated on typical embedded devices, including comparisons with implementations created using alternative DPWS toolkits.*

1. **Introduction**

The advent of the pervasive presence of computing devices, interconnected to form the IoT, brings forth various challenges. End-users do not typically possess the skills to configure and setup the devices that may be found in smart environments; in large-scale deployments, individually setting up devices is not even feasible. From the perspective of developers and implementers, there is a need for rapid development and deployment, while simultaneously tackling issues of scaling and inherent limitations in terms of resources (CPU, memory, power etc.). These are exacerbated by interoperability concerns, as smart devices are offered in variety of hardware platforms and often operate over heterogeneous networks. Moreover, while existing networking mechanisms are updated and adapted to efficiently handle the vast population of these resource-constrained devices (e.g. IETF's work on 6LoWPAN [1] standards to enable IPv6 connectivity over IEEE 802.15.4 networks), higher level, machine to machine interactions, are often required to make use of the devices' full potential. The above have compelled researchers and developers alike to focus on mechanisms that guarantee interoperability, providing seamless access to the various devices and their functional elements. SOAs evolved from this need to have interoperable, cross-platform, cross-domain and network-agnostic access to devices and their services. This approach has already been successful in business environments, as SOAs allow stakeholders to focus on the services themselves, rather than the underlying hardware and network technologies. The DPWS [2] specification enables the adoption of a SOA-based approach on embedded devices with limited resources, allowing system owners to enjoy these benefits across the heterogeneous systems of the IoT ecosystem.

This work presents Node.DPWS, a novel implementation of the DPWS specification using the Node.js platform (http://nodejs.org/), also referred to as Node, a JavaScript-based runtime environment designed to maximize throughput and efficiency. The Node.DPWS libraries leverage the benefits of both DPWS and Node, allowing the creation of high performance, scalable and lightweight DPWS devices on the heterogeneous, often resource-constrained, platforms typically found in smart environments. Moreover, the libraries are easy to use and the devices can be defined using a minimum of code, reducing the development effort.

The article begins with an introduction to DPWS, including its key characteristics and advantages. The related works section (sidebar) details the profile's current status in the industry and research communities and also lists all available alternatives to Node.DPWS (i.e. other libraries and toolkits that can be used to implement DPWS devices). Details about Node.DPWS follow, with a brief introduction to Node.js and a description of the presented DPWS libraries and their features, also providing sample code to demonstrate the implementation approach. The performance of Node.DPWS is evaluated on actual embedded platforms, including a comparison to identical devices developed using the most mature DPWS toolkit currently available. The article concludes with some pointers to future enhancements to Node.DPWS.

## 2. The Devices Profile for Web Services (DPWS)

DPWS was introduced in 2004 by a consortium led by Microsoft and is now an OASIS open standard (at version 1.1 since July 2009). The DPWS specification defines a minimal set of implementation constraints to enable Web Service messaging on resource-constrained devices. It employs similar messaging mechanisms as the Web Services Architecture (WSA, [3]), with restrictions to complexity and message size, allowing the provision of totally platform- and language-neutral services, similar to those offered by traditional web services. As it is built around the OASIS Web Services (WS-) stack, it features superior scaling and seamless integration capabilities, both in the context of enterprise environments (e.g. plant floors or enterprise-wide information systems) and the Internet (e.g. regular W3C-specified Web Services). The SOA-based approach of DPWS can act as an enabler, by providing real-time monitoring and control features usable over the whole smart infrastructure. Allowing the migration of low level information (e.g. sensing data) into higher level contexts (e.g. business operations or knowledge extraction), new types of services are made possible; these are expected to be vital for future end-user as well as enterprise deployments, in a number of industry domains [4].

The profile's architecture includes hosting and hosted services. A single hosting service is associated with each device while the same device may accommodate various hosted services. The latter represent the device's various functional elements and rely on the hosting service for discovery. Thus, a multifunctional device integrated into e.g. a smart home or an enterprise environment, will have a single hosting service but may feature various hosted services (e.g. a temperature service, a door control service, a movement-sensing service etc.). As discovery services are included, the device can advertise its presence on the network and search for other devices. Metadata exchange mechanisms provide dynamic access to services hosted on a device. Additionally, publish and subscribe eventing mechanisms allow clients to subscribe to services provided by devices.

DPWS was originally conceived and introduced as a successor to Universal Plug and Play (UPnP), but due to the lack of backward compatibility such a transition has not taken place yet. Instead, nowadays DPWS is actively pushed by industry stakeholders as the solution of choice for large-scale deployments, while UPnP is mostly targeted to home environments (printers, home entertainment etc.) [5]. Also, like UPnP, DPWS is natively integrated into the various versions of the Windows operating system, from Windows Vista onwards.

The main disadvantage of DPWS, at its current state, is the existence of numerous specifications (protocols, bindings etc.) which have not been consolidated yet.

### 3. Background Material

The deployment and orchestration of web services on heterogeneous embedded devices is a very active research area, following the success of some early research efforts, like the "Service Infrastructure for Real time Embedded Networked Applications" (SIRENA, ITEA2) project which proved the feasibility and advantages of integrating web services, via DPWS, across business segments, including critical sections (e.g. the production floor [6]). Its follow-up projects, "Service-Oriented Device & Delivery Architecture" (SODA, ITEA2) and "Service-Oriented Cross-layer Infrastructure for Distributed Smart Embedded Devices" (SOCRADES, FP6), built upon the work of SIRENA and focused on providing a secure scalable ecosystem and adding more sophisticated features into the SOA-enabled devices to serve the requirements of future manufacturing [7]. More recent research efforts include the "ArchitecturE for Service-Oriented Process" (IMC-AESOP, FP7) and "Web of Objects" (WoO, ITEA2) projects. Researchers in AESOP proposed a system-of-systems approach for monitoring and control, based on a SOA for very large scale (i.e. up to tens of thousands of devices) distributed systems in process control applications. WoO promoted the use of DPWS to build a secure, context-aware service infrastructure for smart objects, focusing on the interoperability of devices & services through the use of semantics.

As many industry leaders from a variety of sectors (electronics, power, automation, enterprise, home etc.) have been involved in the above research efforts, it is evident that there is significant industry backing of DPWS and its use in future applications and products. Moreover, the use and benefits of DPWS have been studied extensively by researchers in the context of various applications areas, which, among others, include automotive and railway systems [8], industrial automation [9], eHealth [10], smart cities [11], smart homes [12] and wireless sensor networks [13]. All of the above are promising indicators for the future of the specification and its wider adoption.

### 3.1. DPWS implementations

A survey on alternative to Node.DPWS APIs for DPWS development reveals a plethora of available solutions with diverse characteristics: the libraries included in Microsoft's .NET Micro Framework (http://netmf.codeplex.com/releases/view/82448), WS4D's toolkits which include WS4D-uDPWS (http://code.google.com/p/udpws/), WS4D-JMEDS (http://sourceforge.net/projects/ws4d-javame/), WS4D-Axis2 (http://trac.e-technik.uni-rostock.de/projects/ws4d-axis2) and WS4D-gSOAP (https://trac.e-technik.uni-rostock.de/projects/ws4d-gsoap/), and, finally, SOA4D's solutions which include DPWS-Core (https://forge.soa4d.org/projects/dpwscore/) and DPWS4J

(https://forge.soa4d.org/projects/dpws4j/). Information for the identified DPWS implementations is aggregated in Table 1.

*Table 1. Overview of DPWS Toolkits*

|  | **.NET Micro Framework** | **WS4D-uDPWS** | **WS4D-JMEDS** | **WS4D-Axis2** | **WS4D-gSOAP** | **DPWS-Core** | **DPWS4J** |
|---|---|---|---|---|---|---|---|
| **Language** | C# | C | Java | Java (Apache Axis2) | C | C | Java |
| **CPU/OS** | ARM | PC (VM), TelosB, AVR Raven | Java SE, Java CDC/CLDC, Android | Java SE | Linux, Windows, ARM | Linux, Windows | Java SE, Java CDC |
| **DPWS 1.0** | √ | - | √ | √ | √ | √ | √ |
| **DPWS 1.1** | √ | √ | √ | - | Partial | Partial | - |
| **IPv4** | √ | √ | √ | √ | √ | √ | √ |
| **IPv6** | - | √ | √ | - | Partial | √ | - |
| **License** | Apache 2.0 | FreeBSD | EPL | Apache 2.0 | GPL/LGPL | LGPL | LGPL |
| **Active** | Yes | No | Yes | No | No | Yes | No |

Nevertheless, when focusing on key features such as code portability, deployment on heterogeneous platforms, support for IPv6 (necessary in the context of the IoT) and active development and support of the tools, the valid options are actually fewer. With the above considerations in mind, WS4D-JMEDS currently stands out as the most attractive choice. It is the most advanced work of the WS4D initiative, providing a mature and feature-rich platform which is being constantly updated and improved. It is, therefore, the toolkit that will be used as a benchmark for the Node.DPWS implementation presented in this work.

**4. About Node.js**

Node.js is a relatively new platform, introduced in May 2009 (in version v0.10.31, as of August 2014). It is an evented server-side implementation based on Google's V8 JavaScript engine. Both Node and the V8 engine are mostly implemented in C and C++, but Node's wrapper enhances the engine's basic features by allowing server-side deployment of JavaScript programs and the use of various C libraries, system calls, binary data manipulation and request handling. The core concept behind its development was to create the building block for lightweight and scalable servers, providing an evented, non-blocking infrastructure to develop highly concurrent applications.

Node handles network input/output (I/O) operations in an evented, non-blocking fashion, while file I/O operations are handled asynchronously. This differentiates Node from typical implementations which follow the threaded model, where for each new connection a thread is created and which have inherent issues with scaling. In Node, each new connection is only a small heap allocation. Moreover, Node's executing thread cannot be blocked; in situations where this would normally happen (e.g. waiting for data from a remote database), the thread's runtime is utilized to serve other requests. The above result in very fast applications which also scale well, even in the case of resource-constrained devices (i.e. with no multi-core processors or large amounts of memory) like those typically embedded in smart devices. More details on Node's characteristics, as well as sample code, can be found in [14].

Research [15] indicates that while Node.js offers significant benefits in terms of I/O operations' performance and resource utilization, it is not as good at serving large static files. This does not harm its applicability in developing fast, scalable network applications, and it is expected to improve as the platform matures. Moreover, it is not an issue in the context of typical IoT applications, where devices are expected to transmit low level information (e.g. sensing data) and receive commands to invoke operations on their functional elements (e.g. turn ON/OFF, set thresholds, rotate etc.). Finally, some concerns regarding the security of Node.js applications can be attributed to the lack of a stable version (as the project is relatively new) and most can be avoided when following security-conscious programming practices [16]. The platform is not inherently insecure, and thus can safely be used in production-grade deployments.

As Node.js addresses many of the issues with developing applications requiring real-time and lightweight communications, it has quickly gained the support of the developers' community (there is a variety of libraries already available for or compatible with Node) and of major stakeholders in the industry [17], including companies such as Google, Microsoft and Yahoo!. Popular websites such as Wikipedia, LinkedIn, eBay and Microsoft's Azure cloud platform already make use of Node.js, even though it has not reached a stable (1.0) version yet; an indicator that there is a strong demand for Node's features and that its user base will continue to grow over time.

**5. The Node.DPWS libraries**

Node's characteristics are a good match for event-driven DPWS-based web services deployable on embedded devices present in smart enterprise, industrial, domestic and other ubiquitous-computing-enhanced environments. Thus it is an attractive solution for implementing the DPWS specification, which could potentially deliver highly scalable DPWS devices, able to handle many clients concurrently, while having very low resource consumption. Node.DPWS provides such an implementation of the DPWS specification using Node.js.

The developer is responsible for describing the device's attributes, its supported services, operation and events, and the libraries will properly advertise them and match them to requests. A key characteristic of Node.DPWS is its ease of use. Operations can be defined with just a few lines of code, and the developer does not have to deal with any low level aspects of the implementation, such as the communication protocols. To highlight these characteristics, the source code of a DPWS device is provided in Figure 1.

```javascript
var dpws = require('dpws')
var server = dpws.createServer({
device: {
// device attributes
address: 'f7ef0fab-ba1d-4275-9a94-0f051090640f',
types: '_PORT_TYPE_',
metadataVersion: Date.now().toString(),
manufacturer: '_MANUFACTURER_',
modelName: '_MODEL_NAME_',
modelNumber: '_MODEL_NUMBER_',
modelUrl: 'http://example.com/_MODEL_URL',
presentationUrl: 'http://example.com/_PRESENTATION_URL',
friendlyName: '_FRIENDLY_NAME_',
firmwareVersion: '0.0.1',
serialNumber: '12345'
}
})

// create service with identifier
var service = server.createService('_SERVICE_ID_', {
// describe supported input/output types
types: {
// create type 'temperature' with a single integer element
'temperature': 'int'
}
})
var temp = 0

// create operation GetTemperature with a single output of type temperature
service.createOperation('GetTemperature', {
output: 'temperature'
}, function (input, cb) {
// cb = callback to call when we get the result.
// call cb without error (null) and temp as the output value.
cb(null, temp)
})

// listen for incoming connections on port 8080
server.listen(8080)
// on CTRL+C, send bye message and exit
process.on('SIGINT', function () {
server.bye(process.exit)
})
```

*Figure 1. Creating a simple DPWS device using Node.DPWS*

In this example Node.DPWS is used to expose a simple service which provides clients with temperature readings (e.g. ones provided by a smart sensor). This temperature operation is defined in a few lines of code; input/output types of the operation are specified and then a handler is provided. The latter will be called whenever an invocation to this operation is made. More complex operations can also be developed in an equally simple manner. In the example of the smart sensor above, a more complex operation could allow clients to subscribe to get temperature readings at set intervals or when certain events occur (following the WS-Eventing specification [18]) and even control the device (setting temperature thresholds, turning it off etc.).

In the proposed solution, nodes support auto-discovery by implementing WS-Discovery [19] which defines a multicast discovery protocol to locate services, with the main mode of discovery being a client looking for one or more target services. Apart from discovering devices, the developed library allows for answering to discovery requests. The developer must provide the details of her device, which the library will then forward to requesting nodes whose queries match the device. The library supports both DPWS version 1.0 (2006) and 1.1 (2009).

**5.1. Performance evaluation**

In order to assess the performance of Node.DPWS and how it compares to alternative solutions, we focused on creating and examining the behavior of a simple DPWS device featuring a "GetTemperature" operation which, when invoked, returned an integer value. Three versions of the device were developed: the Node.DPWS one and two versions using WS4D-JMEDS (i.e. the most appealing alternative implementation currently available), one compiled using Java Standard Edition (SE) and the other following the Java Connected Device Configuration (CDC), which is part of the Java Platform Micro Edition (Java ME) and designed for handheld and embedded systems (i.e. by design better-suited to resource-constrained devices).

The devices were deployed on Beaglebone (http://beagleboard.org/bone) platforms, equipped with a 720MHz ARM Cortex-A8 processor, 256MB of RAM and the Arch Linux ARM (http://archlinuxarm.org/) operating system, interconnected via wired Ethernet, to minimize the effects of the network on the reported performance figures. The test-bed also featured a client application developed to discover and query the DPWS devices, recording response times, for benchmarking purposes.

A total of 500 requests were issued from the benchmarking client (running on a desktop PC) to each of the three DPWS devices, while various aspects of their performance were being monitored and recorded. Figure 2 presents the response time (i.e. the time that the client had to wait before it got a response to its GetTemperature invocation). As is evident from the graph, the Node.DPWS device responded significantly faster than the WS4D-JMEDS-based implementations. Random spikes on the response times can be attributed to the triggering of "housekeeping" operations on the target platforms. Averaging the response times over the 500 requests reveals that each test client got its response from the WS4D-JMEDS SE device after about 72.48ms. The CDC-compiled device responded faster, with an average response time of 51.99ms. Still, the Node.DPWS device performed significantly better than both the WS4D-JMEDS devices, featuring an average response time of 24.44ms, i.e. 66.3% and 53% faster than the SE and CDC variants respectively. Figure 3 better depicts this difference, along with other performance aspects which were observed during benchmarks. The Node.DPWS device was able to reply to twice the number of requests per second than the second best performing device, which is the WS4D-JMEDS CDC one. As a result of the above response time gap, the 500-request benchmark was completed in just 12.22 seconds when querying the Node.DPWS device.

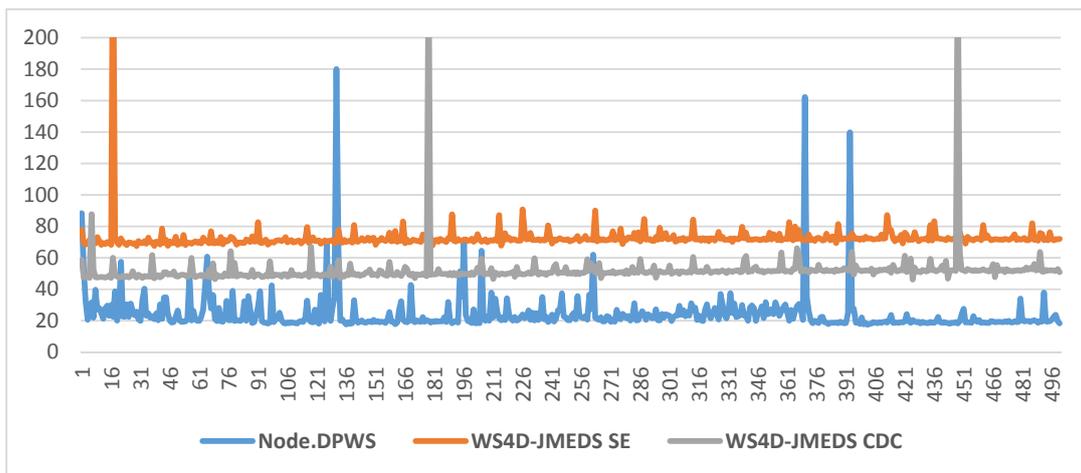

*Figure 2. Response time (in ms) for 500 requests.*

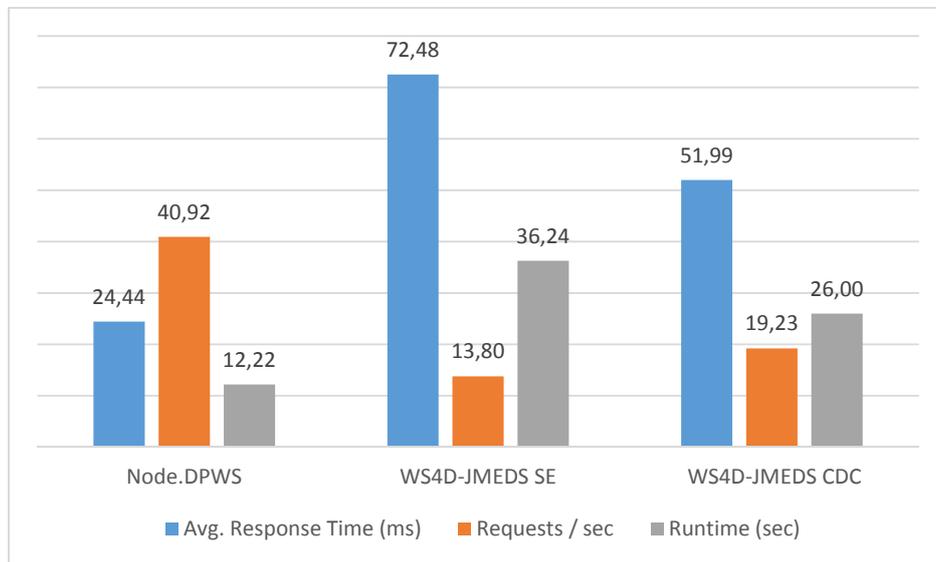

*Figure 3. Performance aspects while handling 500 benchmark requests*

Moving to the performance impact on the DPWS device itself, the load on the ARM processor and memory consumption were recorded during benchmarking. The devices did not have significant differences in terms of CPU load; the average load while handling the test requests was recorded at 90.4%, 91.9% and 96.7% for Node.DPWS, WS4D-JMEDS SE and WS4D-JMEDS CDC respectively. The Node.DPWS device also demonstrated better behavior in terms of the memory consumed during benchmarks. Its memory footprint was measured at 26440 bytes on average, which was 10% lower than its WS4D-JMEDS CDC counterpart (which occupied 29387 bytes) and 18% lower than the WS4D-JMEDS SE one (which consumed an average of 32280 bytes).

## 6. Conclusions

In this article we validated the benefits of leveraging the relatively novel Node.js platform to implement the DPWS specification in the form of Node.DPWS; an easy to use and lightweight set of libraries for creating and deploying DPWS devices on heterogeneous systems with limited resources. The performance assessment revealed that Node.DPWS outperforms the most attractive alternative currently available, the WS4D-JMEDS toolkit. The enhanced performance, scaling and ease-of-development characteristics of Node.DPWS show that there is significant room for improvement in the DPWS-related tools currently available to developers. It is, therefore, worthwhile to pursue further work on the Node.DPWS implementation, in order to enrich its libraries with more features, e.g. to extend the support to other WS-* related protocols, starting with the WS-Security [20] specification. This will also allow us to study the API's behavior in more complex scenarios. Finally, work is under way to provide a website along with proper documentation, to facilitate the use and expansion of Node.DPWS (and the associated Node.js and DPWS technologies) by the research and development community.

## 7. References

[1] G. Mulligan, "The 6LoWPAN architecture," in Proceedings of the 4th workshop on Embedded networked sensors - EmNets '07, 2007, p. 78.
[2] "Devices profile for web services, version 1.1," 2009. [Online]. Available: http://docs.oasis-open.org/ws-dd/dpws/1.1/os/wsdd-dpws-1.1-spec-os.pdf. [Accessed: 23-Aug-2014].
[3] D. Booth, H. Haas, F. McCabe, E. Newcomer, M. Champion, C. Ferris, and D. Orchard, "Web Services Architecture," W3C Working Group, 2004. [Online]. Available: http://www.w3.org/TR/ws-arch/. [Accessed: 23-Aug-2014].
[4] P. Spieß and S. Karnouskos, "Maximizing the business value of networked embedded systems through process-level integration into enterprise software," in 2007 2nd International Conference on Pervasive Computing and Applications, ICPCA'07, 2007, pp. 536–541.
[5] T. Nixon, "UPnP Forum and DPWS Standardization Status," Rally Technologies, Windows Device and Storage Technologies Group, 2008. [Online]. Available: http://download.microsoft.com/download/f/0/5/f05a42ce-575b-4c60-82d6-208d3754b2d6/UPnP_DPWS_RS08.pptx. [Accessed: 23-Aug-2014].

[6] F. Jammes, A. Mensch, and H. Smit, "Service-oriented device communications using the devices profile for Web services," in Proceedings - 21st International Conference on Advanced Information Networking and Applications Workshops/Symposia, AINAW'07, 2007, vol. 2, pp. 947–955.

[7] S. Karnouskos, D. Savio, P. Spiess, D. Guinard, V. Trifa, and O. Baecker, "Real-world Service Interaction with Enterprise Systems in Dynamic Manufacturing Environments," in Artificial Intelligence Techniques for Networked Manufacturing Enterprises Management SE - 14, L. Benyoucef and B. Grabot, Eds. Springer London, 2010, pp. 423–457.

[8] V. Venkatesh, V. Vaithayana, P. Raj, K. Gopalan, and R. Amirtharaj, "A Smart Train Using the DPWS-based Sensor Integration," Res. J. Inf. Technol., vol. 5, no. 3, pp. 352–362, Mar. 2013.

[9] T. Cucinotta, A. Mancina, G. F. Anastasi, G. Lipari, L. Mangeruca, R. Checcozzo, and F. Rusina, "A Real-Time Service-Oriented Architecture for Industrial Automation," IEEE Trans. Ind. Informatics, vol. 5, no. 3, pp. 267–277, Aug. 2009.

[10] S. Pöhlsen, S. Schlichting, M. Strähle, F. Franz, and C. Werner, "A DPWS-Based Architecture for Medical Device Interoperability," in World Congress on Medical Physics and Biomedical Engineering, September 7 - 12, 2009, Munich, Germany SE - 22, vol. 25/5, O. Dössel and W. Schlegel, Eds. Springer Berlin Heidelberg, 2009, pp. 82–85.

[11] K. Fysarakis, K. Rantos, O. Sultatos, C. Manifavas and I. Papaefstathiou, "Policy-based Access Control for DPWS-enabled Ubiquitous Devices," in 19th IEEE International Conference on Emerging Technologies and Factory Automation (ETFA 2014), Barcelona, Spain, Sept. 16-19, 2014.

[12] A. Müller, H. Kinkelin, S. K. Ghai, and G. Carle, "A secure service infrastructure for interconnecting future home networks based on DPWS and XACML," in Proceedings of the 2010 ACM SIGCOMM workshop on Home networks - HomeNets '10, 2010, p. 31.

[13] O. Dohndorf, J. Krüger, H. Krumm, C. Fiehe, A. Litvina, I. Luck, and F. J. Stewing, "Towards the web of things: Using DPWS to bridge isolated OSGi platforms," in 2010 8th IEEE International Conference on Pervasive Computing and Communications Workshops, PERCOM Workshops 2010, 2010, pp. 720–725.

[14] S. Tilkov and S. Vinoski, "Node.js: Using JavaScript to build high-performance network programs," IEEE Internet Comput., vol. 14, pp. 80–83, 2010.

[15] I. K. Chaniotis, K.-I. D. Kyriakou, and N. D. Tselikas, "Is Node.js a viable option for building modern web applications? A performance evaluation study," Computing, Mar. 2014.

[16] A. Ojamaa and D. Karl, "Security Assessment of Node.js Platform," in Information Systems Security, 2012, pp. 35–43.

[17] "Projects, Applications, and Companies Using Node." [Online]. Available: https://github.com/joyent/node/wiki/Projects,-Applications,-and-Companies-Using-Node. [Accessed: 23-Aug-2014].

[18] D. Box, L. F. Cabrera, C. Critchley, F. Curbera, D. Ferguson, S. Graham, D. Hull, G. Kakivaya, A. Lewis, B. Lovering, P. Niblett, D. Orchard, S. Samdarshi, J. Schlimmer, I. Sedukhin, J. Shewchuk, S. Weerawarana, and D. Wortendyke, "Web Services Eventing (WS-Eventing)," W3C Member Submission, vol. 2009. pp. 1–34, 2006.

[19] T. Nixon, A. Regnier, V. Modi, and D. Kemp, "Web Services Dynamic Discovery (WS- Discovery), version 1.1," OASIS Standard Specification, 2009.

[20] K. Lawrence, C. Kaler, A. Nadalin, R. Monzilo, and P. Hallam-Baker, "Web Services Security: SOAP Message Security 1.1," OASIS Standard Specification, 2006.